\documentclass[12pt]{article}
\usepackage{times}
\pdfoutput=1
\usepackage{graphicx}
\usepackage{geometry}
\usepackage[utf8]{inputenc}
\usepackage{ragged2e}
\usepackage{pifont}
\usepackage{wrapfig}

\begin{document}
\raggedright
\huge
Astro2020 Science White Paper \smallskip

The Extended Cool Gas Reservoirs Within $z > $1 (Proto-)Cluster Environments

%\linebreak

\bigskip

\normalsize

\noindent \textbf{Primary Thematic Area:} Galaxy Evolution 

\textbf{Principal Author:}

Name:\,Kevin~C.~Harrington
 \linebreak						
Institution:\, \textit{Argelander Institut f\"{u}r Astronomie, Auf dem H\"{u}gel 71, 53121 Bonn,Germany}\\
\textit{Max-Planck-Institut f\"{u}r Radioastronomie, Auf dem H\"{u}gel 69, 53121 Bonn, Germany} \\
Email:\,kharring@astro.uni-bonn.de
 \linebreak
Phone:+49 - 228 -73 - 3521

\vspace{-4pt}
\justify        
        
\textbf{Co-authors:~}\\
David Frayer \\
\textit{Green Bank Observatory, 155 Observatory Rd., Green Bank, West Virginia 24944, USA}\\
\linebreak
Helmut Dannerbauer \\
\textit{Instituto de Astrof{\'i}sica de Canarias (IAC), E-38205 La Laguna, Tenerife, Spain\\
Universidad de La Laguna, Dpto. Astrof{\'i}sica, E-38206 La Laguna, Tenerife, Spain}\\

%\clearpage\maketitle
%\thispagestyle{empty}
       \textbf{Abstract:} High-redshift ($z$) proto-clusters will serve as testing grounds to probe the gas supply furnishing the emerging metals, stars, and large-scale structures we see at the current epoch. This work focuses on the major role large radio/millimeter (mm) single dish facilities will have in constraining the bulk, cold (T $= 10^{1-4}$K) molecular and atomic gas content. To highlight the need for large radio/mm single dishes, we calculate how the high-sensitivity of the Green Bank Telescope's (GBT) unblocked 100m aperture provides vital interferometric short-spacing coverage to support higher-resolution ngVLA observations of the cold neutral gas at the largest scales. These combined observations are optimal for revealing low-surface brightness emission, and thus aid in the total baryonic mass estimates across cosmic time.

\pagebreak

%\lipsum[1-10]
%\pagenumbering{num_style}
\textbf{1. Introduction}
% {\bf A key missing-link in our understanding of the formation and evolution of galaxies is how the gas evolves within proto-cluster environments at high redshift to form the galaxies, stars, and metals that we see at the current epoch.} 

%DF; 
%In the previous decade, observations with ALMA and the VLA have made enormous strides in studying high-redshift starbursts and AGN with high-resolution imaging, uncovering young gas-rich disks (Refs Tacconi, Daddi,.. Genzel (or whoever)), gas-rich mergers (refs), evidence for AGN feedback, ect.   

%In order to trace the build-up of cold neutral gas across cosmic time, future observations will require sampling wide-areas of the sky to overcome cosmic variance and to trace the total cold neutral gas content. The next decade will shed light on the physical processes involved with the rapid gas fueling of massive systems between the epoch of reionisation ( $z \sim 6.5-9$ ), and the associated low-metallicity gas streams within the IGM. These processes lead to the first enrichment of metals and, in turn, drives the increased gas-to-stellar mass fractions for the actively star-forming galaxies at $z \sim 1-4$ \cite{Tacconi2018}.\\
\begin{figure*}[!b]
\centering
\includegraphics[width=0.8\textwidth]{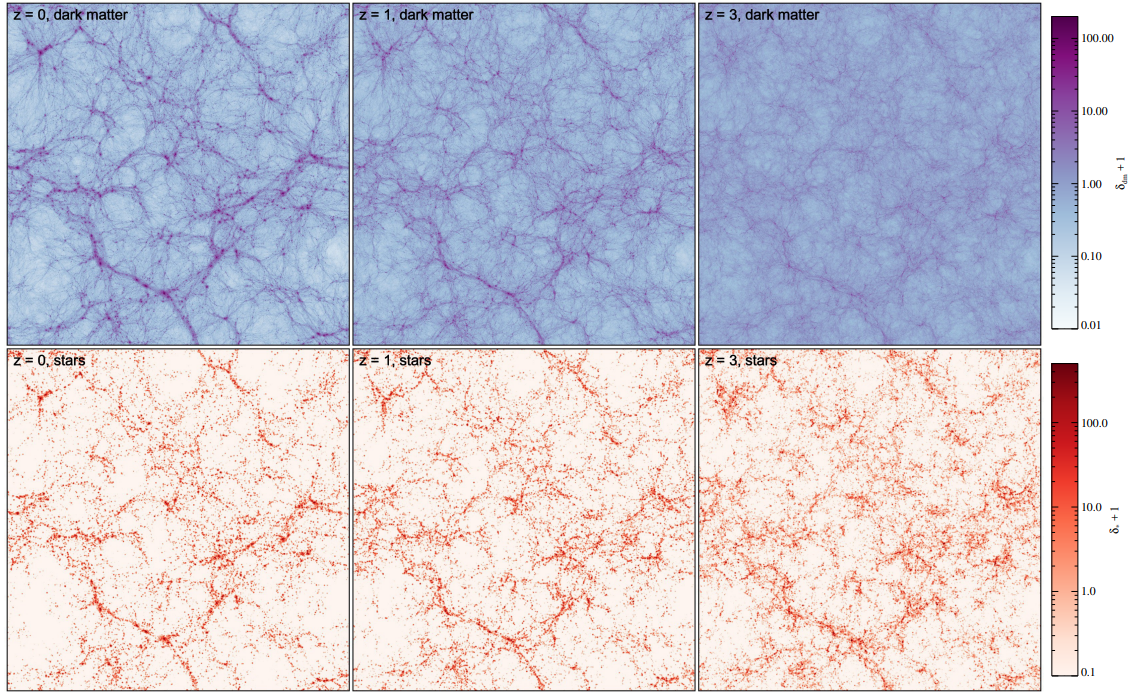}
\caption{First results from the IllustrisTNG (TNG300); \cite{springel18}. \textbf{Top:} From left to right, the dark matter distribution at $z = $0, 1, 3. The color bar range for $\delta_{\rm DM} + 1$ is 0.01 to 100. \textbf{Bottom:} From left to right, the stellar mass distribution at $z = $0, 1, 3. The color bar range for $\delta_{\rm \star} + 1$ is 0.1 to 100.  The baryonic (dark matter) mass resolution is 7.44$\times 10^{6} h^{-1}$ (3.98$\times 10^{6} h^{-1}$) M$_{\rm \odot}$, with a box side length of $205 h^{-1}$ Mpc and thickness of $25 h^{-1}$ Mpc }
\end{figure*}

The multi-phase gas distribution within/surrounding proto-cluster environments requires observations that will transform our understanding of how high-redshift (i.e. $z > 1$) galaxies form within local cluster and filamentary structures. The rapid development in space-based IR/sub-mm facilities, such as the Origins Space Telescope (\textit{Origins}), will yield the most important spectral line diagnostics of the warm and dense star-forming gas (T $>$ 100 K; $n_{\rm H_{2}} >$ 10$^{4}$). In order to probe the cooler, and more diffuse, cosmic gas supply, future ground-based efforts in the mm/radio will be required.\\

In the previous decade, cool molecular/atomic gas studies at high-$z$ have concentrated on individual Active Galactic Nuclei (AGN) or massive star-forming systems, with bright infrared (IR) luminosity, i.e. $L_{\rm IR} > 10^{12.5}$ L$_{\rm \odot}$ \cite{carilli13}.  Currently, many high-$z$ systems have low/mid-J transitions, tracing primarily molecular gas associated with ongoing SF, resulting in bulk gas estimates sensitive to the uncertainties in the scaling relations developed for field galaxies \cite{Tacconi2018}. Recently there has been an increase in high-resolution imaging of cool gas with the VLA and ALMA, while the cold gas supply from the CGM/IGM that is fundamentally responsible for fueling the rapid stellar-mass growth and metal enrichment has yet to be mapped systematically in galaxy proto-clusters. Technological advancements in the coming decade will overcome the challenges in observing this low-surface brightness emission across the largest spatial scales, yielding the required comparisons to detailed cosmological model predictions \cite{hayward13,sparre15,sparre17}.\\ 
Star-forming galaxies mostly form in groups or clusters within massive dark matter haloes $\sim 10^{11-13} {\rm M_{\odot}}$, and the clustering of baryonic gas is predicted to be stronger than the dark matter distribution at $z\sim 3$ \cite{springel18} (IllustrisTNG; Fig.1). In particular, observations of the large-scale distribution of cool neutral gas within the circumgalactic or intergalactic media (CGM;IGM) at high-$z$ will reveal \textbf{(i.)} the fueling processes involved in the rapid stellar mass growth at $z \sim$ 3-6 which led to $\sim$50\% of the total galaxy population at $z \sim 2$ having already quenched their SF activity \cite{Toft2014SubmillimeterGalaxies} and \textbf{(ii.)} the total bulk molecular gas mass and role of IGM gas accretion in fueling the co-moving SFR density peak at $z \sim$ 1-3. \\
\begin{figure*}
%\begin{wrapfigure}[10]{o}{0.495\textwidth}
  \centering
    \vspace{-.02pt}
    \includegraphics[width=0.485\textwidth]{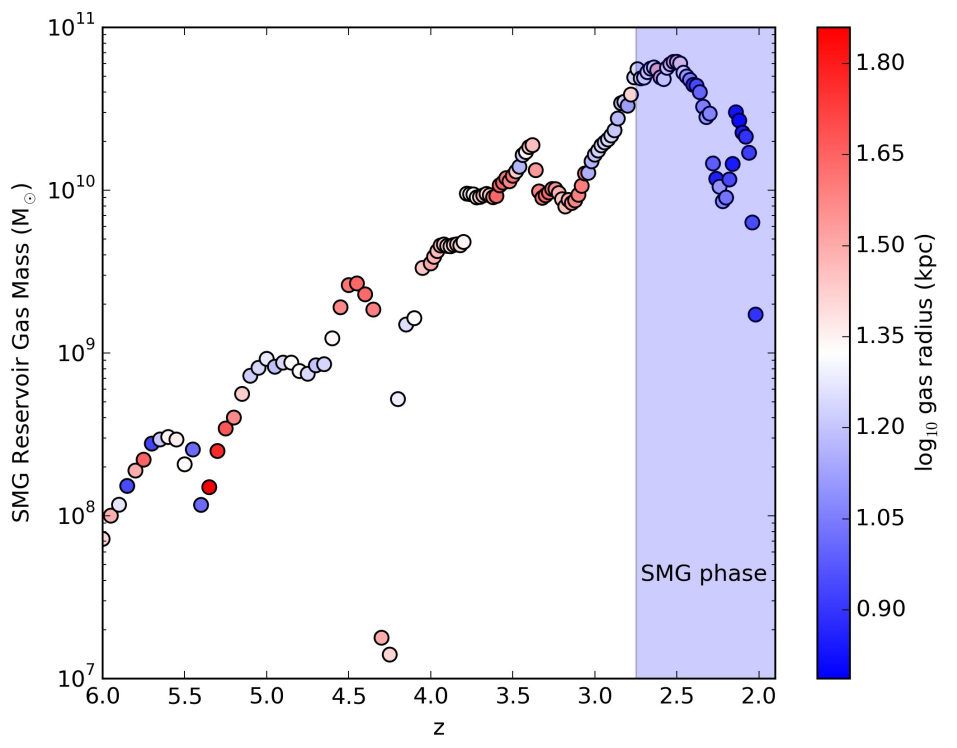}
    \caption{From Narayanan et al. 2015; \cite{narayanan15}- Theoretical molecular gas consumption in central galaxy during a dusty starburst phase, as predicted for a bright submillimeter galaxy (SMG). Colors mark the median scale height from the center of mass in the galaxy.}
    \label{sample}
\end{figure*}
%\end{wrapfigure}

%\centering
%\includegraphics[scale=0.19]{frayer-propfig}
%\includegraphics[scale=0.19]{narayanan15}
%\vspace{-0.03pt}
%\caption{\textbf{Left:} The SSA22 field showing the Lyα emission objects (LAEs) and the SMGs (red dots) as function of redshift and location within the field \cite{matsuda05,umehata15}. The green line shows the filamentary structure defined by the overdensity of LAEs. The rectangular region within the orange dotted lines (lower-right panel) shows the location of the ALMA Deep field in SSA22 \cite{umehata15}.\textbf{Right:} \cite{narayanan15}-- \textit{ "Mass of gas in central galaxy that will be consumed during SMG starburst as a function of z. The colours denote the median scale height from the galaxy centre of mass."}}
%\end{figure*}

%\textbf{Bottom:}\cite{narayanan15} show a 250 kpc-sized surface density projection map at $z = 3.0, 2.8, 2.6$. Green, yellow and red colors correspond to 10$^{17-19}$ N$_{H}$g cm$^{-2}$. }

%Our focus is complementary to \cite{emonts18ngvla}, who have highlighted the advantages of the planned, core-configuration of the ngVLA to access larger spatial scales to research the low-surface brightness emission from extended cold gas in the circumgalactic or intergalactic medium (CGM;IGM). 

\textbf{2. Gas Rich Environs of Proto-clusters}\\

%Within the last five to ten years, the study of high-z proto-clusters has steadily evolved.  \textit{Tracing the overall gas content and excitation conditions that are embedded within and surrounding gas-rich proto-clusters environs in the $z > 1$ CGM and IGM is a central objective in the coming decade. }
The coalescence of proto-clusters occurs during the peak epoch of the co-moving star-formation rate (SFR) density at $1 < z < 3$ \cite{overzier16}. Although observations of the cold gas supply from the IGM has not been fully explored, galaxy proto-clusters are likely responsible for explaining the massive-end of the red sequence of galaxies with quenched SF within massive galaxy clusters at low-$z$ \cite{bell04}. Overdense fields at high-$z$ can exhibit a diverse clustering of Ly-$\alpha$ emitters (LAEs) out to 10s of Mpc, with dusty star-forming galaxies clustering at the cores of such systems \cite{matsuda05,umehata15}. In addition, low surface-brightness gas may form as the CGM/IGM cools outside of the hot perimeter of the extended Ly-$\alpha$ emission ($>$ 10s-100 kpc) observed in QSOs and strong LAEs at $z\sim 2-3$ \cite{ArrigoniBattaia2018QSOQuasars,cai17}.\\

A large number of massive over-dense regions at z$>$1 have been identified using the all-sky Planck colors \cite{planckcollab15overdensities, clements14}, while the past five to ten years has seen a growing number of proto-cluster studies \cite{Aravena2012DeepZ=1.5,tadaki14,hayashi17,noble17,stach17,rudnick17,Lee2017,coogan18,hayashi18}. One such example includes the radio-selected $z = $2 proto-cluster, the Spiderweb \cite{danner14,emonts16a,emonts18,Gullberg2016ALMAWeb}, within which individual proto-cluster galaxies have a velocity dispersion, $\sigma_{galaxy} \sim1000 $km s$^{-1}$, and cold CO (1-0)-emitting gas  spread across more than 50 kpc (with $\sigma_{CO} \sim 200 $km s$^{-1}$). Observations at $z \sim$ 4 reveal several starbursting galaxies which dominate the rapid stellar mass assembly ( L$_{\rm IR} > 10^{14} {\rm L_{\odot}}$) of a concentrated proto-cluster environment that encompasses 140-280 kpc  \cite{oteo18, miller18}. Depending on the available gas in the IGM, this galaxy may proceed towards being one of the most massive structures of the local Universe.\\ 

Most recently, \cite{casey18ngvla} have emphasized the importance of CO(1-0) to trace the total molecular gas, as even the CO(2-1), and certainly the CO(3-2), line emission can begin to trace spatially distinct, and denser gas regions (also see e.g. Ivison et al. 2011, Oteo et al. 2016, 2017). This ultimately biases dynamical mass estimates. Theoretically, both the atomic carbon and CO(1-0) trace similar volumes within gas clouds, and recently the atomic carbon line has been developed as a total gas mass tracer \cite{Papadopoulos2004,Weiss2005,glover16}. Since CO(1-0) has an energy requirement of 5.5K above ground, the gas in the CGM can also be excited by the CMB at higher redshift if not shielded, making multiple low-J and [CI](1-0) emission line maps useful to measure the CMB effects \cite{zhang16, daCunha2015} when viewed on the largest scales in-between galaxies. This is of particular importance for molecular gas that eventually breaks down into atomic form when ejected/stripped from a galaxy \cite{leroy15}. \\
Fig. 2 Right, adopted from \cite{narayanan15}, shows the strong rise, with increasing redshift, of the gas consumption-to-stellar-mass assembly of a galaxy, excluding the gas that may escape the system into the CGM/IGM. The relative scale height for the molecular gas, before it is re-accreted, oscillates as the redshift increases until the peak epoch of co-moving SFR density. In these cosmological, hydrodynamical zoom simulations there is a one to two order of magnitude increase in molecular gas mass consumed by a typical dusty, star-forming galaxy between $z \sim$ 5.5- 2.5 \cite{narayanan15}, suggesting that the increased molecular gas mass that is processed by an individual galaxy is fueled by the available supply from the in-flowing gas at distances much greater than the scale heights of the measured central galaxy ($>> 10-100$ kpc). The most massive growth events for a proto-cluster, occuring at $z \sim$ 5.5- 2.5 (e.g. Chiang et al. 2017), mark the most dynamic interplay between baryonic cooling, dark matter collapse, and the hierarchical growth processes across cosmic time. \\

\textbf{3. The Need for Large radio/mm Single-Dishes in the Next Decade}\\

The 100m, unblocked aperture Green Bank Telescope (GBT) will serve as a leading single-dish facility in the coming decade, providing the essential interferometric short-spacing, swift mapping speeds and consistent sky frequency coverage as the VLA/ngVLA to detect the low-surface brightness emission from extended cold gas.\\

Large single-dishes complement interferometers by enabling science on spatial scales that are resolved out by interferometers, and the GBT is the only facility currently operating over the full range of proposed ngVLA frequencies (1-116 GHz). The previous decade has recently outgrown the era of limited bandwidth\footnote{Spectrometers used to have equivalent velocity coverage of $\le \pm$1000 km/s, i.e. only encompassing the total emission line, and could therefore miss a significant amount of flux for a FWZI of 750 km/s or greater (i.e. for $\nu < 250 $GHz;receiver bandwidths are even smaller for higher frequency).}, and the door is now open to investigate the total gas contents of merger or cluster environments with atomic/molecular emission line profiles having full width at zero-intensity (FWZI) $\ge$ 1000 km/s \cite{harris12}. Large volumes of redshift-space can be observed with increased spectrometer bandwidth capability. The GBT will specifically be able to systematically map the redshifted atomic carbon [CI](1-0), CO(1-0; 2-1; 3-2) line emission surrounding the most massive, over-dense regions of gas-rich star-forming systems between about 3.3 $< z <$ 5.6. In doing so, the GBT will also be able to search for the existence of previously undetected, low-excitation, gas-rich systems with dimmer $L_{\rm IR}$ than dusty star-forming systems, yet with comparable gas mass (see remark by Carilli \& Blain, 2002).\\

%The GBO has always been a strong instrumentation site, with current plans that target the development and commissioning of new array cameras from $\lambda \sim$3mm down to the lowest frequencies available for the VLA/ngVLA. Future instrumentation is expected to develop multi-pixel spectrometres at 18-50 GHz, with fast mapping speed and functionality as seen in the latest ARGUS (and planned ARGUS$^{+}$) spectral line imaging at 74-115 GHz.

\begin{figure*}[]
\centering
\includegraphics[width=0.8\textwidth]{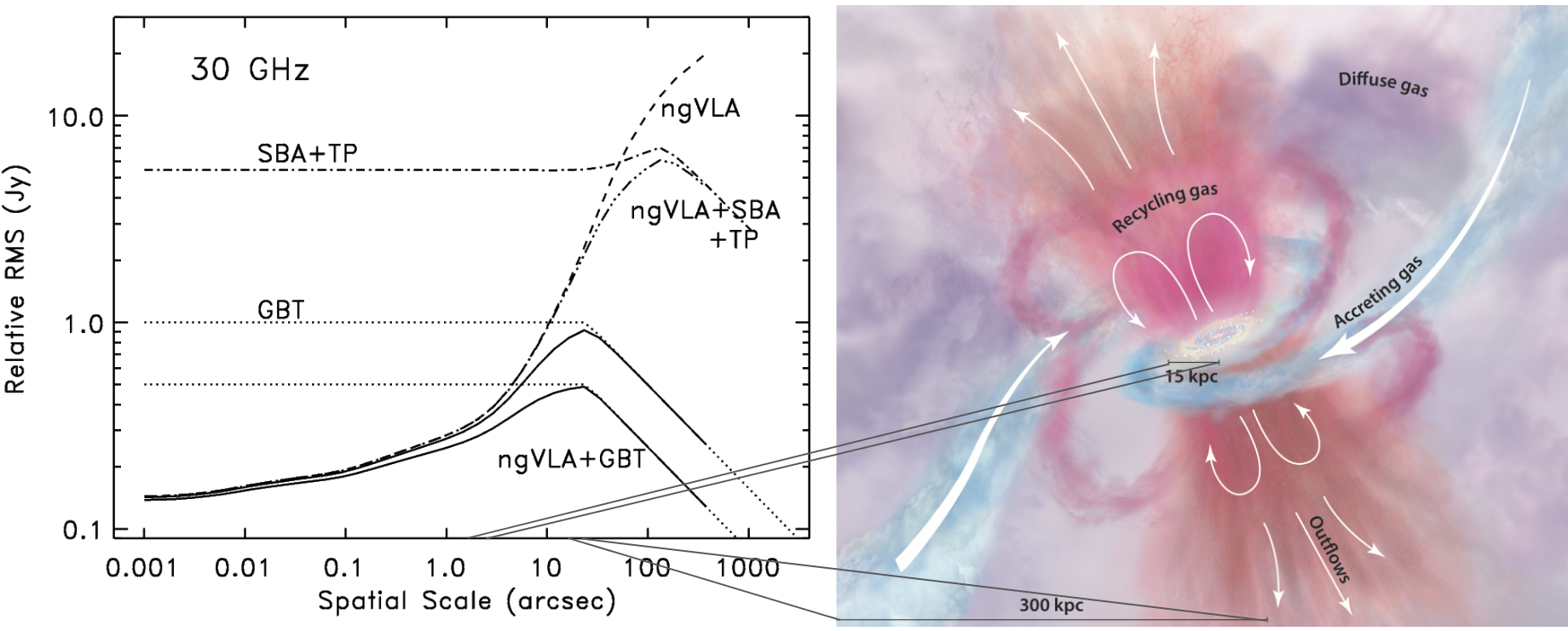}
\vspace{-0.03pt}
\caption{\textbf{Left:} Relative sensitivity in flux density at 30 GHz for the same amount of integration time as a function of spatial scale: Both dotted lines show the sensitivity of the GBT with current technology (upper dotted) and with the expected improvement of a factor of two with future background-limited bolometer spectrometers (lower dotted line); the ngVLA (dashed line) has excellent sensitivity on small scales, but poor sensitivity on large scales even with the inclusion of the planned ngVLA SBA+TP data (dashed-dotted line). The combination of GBT and ngVLA data (solid lines) provides the best sensitivity over all spatial scales. \textbf{Right:}  The relevant physical scales are shown with respect to the spatial scales on the left; \cite{tumlinson17} \textit{ "A cartoon view of the CGM."}}
\end{figure*}

\textbf{4. The Combination of the GBT with the ngVLA}\\

To highlight the importance of the GBT for the future of radio astronomy, we estimate the relative sensitivity of the GBT compared to the proposed ngVLA as a function of spatial scale at 30 GHz, corresponding to redshifted CO(1-0) emission at $z=2.8$ (Fig. 3).   The sensitivity of the ngVLA visibilities were computed using the appropriate weights associated with tapering the data to match the spatial scale (ngVLA Memo\#14;\cite{frayer17}), assuming the updated 2018 ngVLA reference design (214 x 18m dishes). The sensitivity of the ngVLA falls off exponentially for sources larger than about 1 arcsec. The ngVLA project plans to provide short-spacing data by using a Short Baseline Array (SBA) comprised of 19 6m dishes and four 18m dishes in total-power mode (TP). The combination of the SBA+TP data from the ngVLA project would be nearly an order of magnitude less sensitivity than GBT with current technology, and more than a than an order of magnitude less sensitive using future background-limited bolometer spectrometers that can be deployed only on single-dishes (e.g., Branford et al. 2008, SPIE, vol 7020, 70201O with TES bolometer technology and/or potentially using the newer MKID bolometer technology). By itself, the ngVLA will only be useful for observing emission on the smallest spatial scales. To study larger spatial scales, e.g., $>\sim10$ arcsec ($>75$ kpc at $z=$2--3), the GBT (or a similarly sized single-dish) would be needed, in addition to the $\sim 6"$ ATCA beam-size (Emonts et al. 2018). The combination data from the GBT and the ngVLA would provide the highest envisioned sensitivity over all spatial scales (Fig. 3). With the ngVLA+GBT, we will be able to study galaxy formation over spatial scales ranging from 10s of pc to more than 100 Mpc.\\

%\textit{Side-note} -- the more sensitive bolometer technology cannot be deployed on interferometers since the phase of the visibilities (which is not preserved with bolometers) is required for interferometric techniques. Only single-dishes can make use of the advantages provided by the new background-limited wide bandwidth spectrometers enabled with bolometer technologies.
\textbf{5. Outstanding Questions and Outlook for 2020-2030}\\

The combination of the GBT, and other large single dish radio/mm facilities, with the ngVLA will provide a powerful instrument capable of addressing the following open questions (in no particular order):
\newline

$\bullet$ How does the gas depletion time, i.e. the ratio of the total molecular gas mass to star-formation rate (SFR), change as a function of spatial proximity to a proto-cluster core?  
\newline

$\bullet$ A possible gradient in the excitation conditions of the gas in a diverse population of galaxies (e.g. an overdensity of both LAEs and SMGs) within co-moving volumes on the order of 10s of Mpc has been largely unexplored. How would the gas excitation conditions differ within the interstellar medium of a high-$z$ cluster member vs. the global cluster-scale excitation? And, how is this (un)affected by the variations in these diagnostics across the CGM/IGM of a massive gas-rich, proto-cluster environment? 
\newline

$\bullet$ How does the brightness temperature ratios of low-density gas tracers (n$_{\rm H_{2}} < 10^{1-3}$ cm$^{-3}$), i.e. carbon and low-J CO, depend on the CGM ($> 10s-100s$ kpc) of a QSO versus a SFG? \\
\newline

$\bullet$ How enriched (e.g. X([CI])/X(CO) abundances) is the CGM/IGM with respect to a field or cluster galaxy, and to what extent can a cloud self-shield itself in the CGM/IGM? It is likely that carbon and CO abundances would diminish with growing distances from proto-cluster core, but can dense gas that is stripped during galaxy mergers retain its composition before being captured by a gas stream or galaxy in the CGM/IGM? What processes play a dominant role in the increased ram-pressure stripping \cite{bekki09,ebeling14,darvish18}?\\
\newline

$\bullet$ The low-excitation gas traced by mm/radio facilities at 0.1-10s Mpc scale will aid large field of view observations in the  IR/sub-mm (e.g. \textit{Origins}) of the more highly excited gas (including SF/AGN feedback) in the CGM/IGM at $z > 1$. \\
\newline

%$\bullet$Will galaxy growth and increased SF within massive dark matter halos increase the halo temperature such that the gas from the major gas-rich system will cool within the cooler, satellite systems of lower stellar mass? What are the physical gas properties of proto-cluster fields that influences the efficiency of gas accretion for an individual proto-cluster member galaxy \cite{dekel09}? Or in other words, at what point will a galaxy no longer retain the fresh supply of gas due to cluster disturbances? \newline
Other Astro2020 science white papers (Casey et al. 2018; Emonts et al. 2018) also highlight the importance of this area of research. Strong progress and conclusive observations will be made in the coming decade with regards to the thermodynamic atomic/molecular gas properties within and surrounding the $z > 1$ CGM/IGM. Studying the total gas content of the rapidly forming progenitors of clusters and filaments of galaxies at $z\sim 0$ requires sensitive single-dish observations. These necessary measurements will then place the tightest dynamical mass constraints on the baryonic fraction in dark matter halos. Interferometers, by themselves, cannot study the distribution of gas on large spatial scales because the low-surface brightness emission will be resolved out. Therefore single-dish facilities will continue to advance our understanding of the coolest atomic/molecular gas from the interstellar to intergalactic territory at $z > 1$. 

\pagebreak
%%%\bibliography{bigbibs.bib}
%%%\bibliographystyle{jponew.bst}
{}

\end{document}